\documentclass[bm,cite,figures]{epl}
\usepackage{psfrag}
\usepackage{graphicx}
\usepackage{amssymb,amsmath}

\title{Breakdown of {G}allavotti-{C}ohen symmetry for stochastic dynamics}
\shorttitle{Breakdown of {G}allavotti-{C}ohen symmetry ...}
\author{R.~J. Harris\inst{1}\thanks{E-mail: \email{r.harris@fz-juelich.de}} \and A.~R{\'a}kos\inst{2} \and G.~M. Sch{\"u}tz\inst{1}}
\shortauthor{R.~J. Harris \etal}
\institute{
\inst{1} Institut f\"ur Festk\"orperforschung, Forschungszentrum J\"ulich - 52425 J\"ulich, Germany \\
\inst{2} Department of Physics of Complex Systems, Weizmann Institute of Science - Rehovot 76100, Israel
}


\begin{document}

\maketitle

\begin{abstract}
We consider the behaviour of current fluctuations in the one-dimensional partially asymmetric zero-range process with open boundaries.  Significantly, we find that the distribution of large current fluctuations does not satisfy the Gallavotti-Cohen symmetry and that such a breakdown can generally occur in systems with unbounded state space.  We also discuss the dependence of the asymptotic current distribution on the initial state of the system.

\end{abstract}


Substantial progress in the understanding of nonequilibrium systems has been achieved recently through so-called fluctuation theorems~\cite{Evans02b}.
Specifically, the Gallavotti-Cohen fluctuation theorem (GCFT) can be loosely written as
\begin{equation}
\frac{p(-\sigma,t)}{p(\sigma,t)} \sim  e^{-\sigma t} \label{e:GCFT}
\end{equation}
where  $p(\sigma,t)$ is the probability to  observe an average value $\sigma$ for the entropy production in time interval $t$ and $\sim$ denotes the limiting behaviour for large $t$. This theorem was first derived for deterministic systems~\cite{Gallavotti95} (motivated by computer simulations of sheared fluids~\cite{Evans93}) and subsequently for stochastic dynamics~\cite{Kurchan98, Lebowitz99}.  From~\cite{Ciliberto98} onwards there have been successful attempts at experimental verification, including for simple random processes such as the driven two-level system in~\cite{Schuler05}.  Strictly the GCFT is a property of non-equilibrium steady states but, for systems with a unique stationary state it is usually also expected to hold for arbitrary initial states (see, e.g.,~\cite{Cohen99,Searles99} for discussion on this point).  We will refer to this more general property of the large deviation function as ``GC symmetry''.  Some related issues have previously been discussed for Langevin dynamics~\cite{Kurchan98,Farago02,Baiesi06}; we consider the more general case of stochastic \emph{many-particle} systems.

Specifically, we explore the GC symmetry in the context of a paradigmatic non-equilibrium model---the zero-range process~\cite{Spitzer70}.  For certain parameter values, this interacting particle system exhibits a condensation phenomenon~\cite{Evans00,Jeon00b} in which a macroscopic proportion of particles pile up on a single site.  Condensation transitions are well-known in colloidal and granular systems~\cite{Shim04} and also occur in a variety of other physical and nonphysical contexts~\cite{Evans05}.  In~\cite{Me05} it was argued that current fluctuations in the asymmetric zero-range process with open boundary conditions can become spatially-inhomogeneous for large fluctuations---a precursor of the condensation which occurs for strong boundary driving.  Here, for a specialized case, we explicitly calculate the current distribution in this large-fluctuation regime and thus prove a breakdown of the symmetry relation~\eqref{e:GCFT}.  
Significantly, we argue that our analytical approach predicts that this effect also occurs for more general models.
Fianlly, we discuss the relation of our results to GCFT breakdowns found in some other works~\cite{Bonetto05,vanZon03,vanZon04}.


Let us begin by defining our model---the partially asymmetric zero-range process (PAZRP) on an open one-dimensional lattice of $L$ sites~\cite{Levine04c}.  Each site can contain any integer number of particles, the topmost of which hops randomly to a neighbouring site after an exponentially distributed waiting time.  In the bulk particles move to the right (left) with rate $p w_n$ ($q w_n$) where $w_n$ depends only on the occupation number $n$ of the departure site.  Particles are injected onto site 1 ($L$) with rate $\alpha$ ($\delta$) and removed with rate $\gamma w_n$ ($\beta w_n$).  If the partition function has a finite radius of convergence (i.e, $\lim_{n\to\infty} w_n$  is finite) then for strong boundary driving a growing condensate occurs at one or both of the boundary sites~\cite{Levine04c}.  

We are interested in the probability distribution of integrated current $J_l(t)$, i.e., the net number of particle jumps between sites $l$ and $l+1$ in time interval $[0,t]$.  The long-time asymptotic behaviour of this distribution is characterized by the limit of the generating function
\begin{equation}
e_l(\lambda) =\lim_{t \rightarrow \infty} - \frac{1}{t} \ln {\langle e^{-\lambda J_l(t)} \rangle}. \label{e:e_l}
\end{equation}
which implies~\cite{Lebowitz99} a large deviation property for the asymptotic probability distribution, $p_l(j,t)=\mathrm{Prob}(j_l=j,t)$, of the observed ``average'' current $j_l=J_l/t$
\begin{equation}
p_l(j,t) \sim e^{-t\hat{e}_l(j)} \label{e:pj}
\end{equation}
where $\hat{e}_l(j)$ is the Legendre transformation of $e_l(\lambda)$, i.e.,
$
\hat{e}_l(j)=\max_{\lambda}\{e_l(\lambda)-\lambda j \}. \label{e:lang}
$

To calculate the current distribution we employ the quantum Hamiltonian formalism~\cite{Schutz01} where the master equation for the probability vector $|P_t\rangle$ resembles a Schr\"odinger equation with Hamiltonian $H$ (see~\cite{Levine04c} for details).
The generating function $\langle e^{-\lambda J_l(t)} \rangle$ can then be written as $\langle s | e^{-\tilde{H}_l t} |P_0\rangle$ where $\tilde{H}_l$ is a modified Hamiltonian in which the terms in $H$ giving a unit increase/decrease in $J_l$ are multiplied by  $e^{\mp\lambda}$~\cite{Me05}.  Here $|P_0\rangle$ is the initial probability distribution and $\langle s|$ is a summation vector giving the average value over all configurations.  For the current into the system from the left (which can be positive or negative) we consider $\tilde{H}_0$ with lowest eigenvalue $\tilde{e}_{0}(\lambda)$ and corresponding eigenvector $|\tilde{0}\rangle$.  In the case where $\langle s | \tilde{0} \rangle$ and $\langle \tilde{0} | P_0 \rangle$ are finite, the long-time limiting behaviour is given by
\begin{equation}
\langle e^{-\lambda J_0(t)} \rangle \sim \langle s |\tilde{0} \rangle \langle \tilde{0} | P_0\rangle e^{-\tilde{e}_0(\lambda) t} \label{e:pre}
\end{equation}
In this case we have $e_0(\lambda)=\tilde{e}_0(\lambda)$ and the form of $\tilde{H}_0$ imposes the GC symmetry relation
\begin{equation}
e_0(\lambda)=e_0(2E-\lambda) \label{e:GCFTe}
\end{equation}
which leads, via~\eqref{e:pj}, 
to the relationship~\eqref{e:GCFT} with $\sigma=2Ej$ and effective field $E$ given by $e^{2E}= (\alpha \beta / \gamma\delta ) ( p/q )^{L-1}$.  The field $E$ can be related to a force $F=2Ek_BT$.


While the ground-state eigenvalue calculated in~\cite{Me05} is independent of $w_n$, the latter determine the form of the eigenvectors $\langle \tilde{0}|$ and $|\tilde{0} \rangle$.  If $\lim_{n\to\infty} w_n$ is finite then $\langle s | \tilde{0} \rangle$ diverges for some values of $\lambda$.  For a fixed initial particle configuration $\langle \tilde{0} | P_0 \rangle$ is always finite.  However, for a normalized distribution over initial configurations (e.g., the steady-state) $\langle \tilde{0} | P_0 \rangle$ can also diverge (again in the case where $w_n$ is bounded) meaning that \emph{the asymptotic current distribution retains a dependence on the initial state.}  
This has important 
consequences for measurement of the current fluctuations in simulation (or equivalent experiments).  Suppose we start from a fixed initial particle configuration, e.g., the empty lattice, wait for some time $T_1$ and then measure the current over a time interval $T_2$.  These are two noncommuting timescales---if we take $T_2 \to \infty$ faster than $T_1 \to \infty$ we will measure the asymptotic distribution of current fluctuations corresponding to the fixed initial condition which may differ from the asymptotic behaviour of steady-state current fluctuations obtained by taking $T_1 \to \infty$ before $T_2 \to \infty$.


We first specialize to the case of the single-site PAZRP, i.e, one site with ``input'' (left) and ``output'' (right) bonds.  In this model explicit calculation of the matrix element $\langle s | e^{-\tilde{H}_0 t} |P_0 \rangle$ is possible.  For simplicity we consider here $w_n=1$, anticipating qualitatively the same effects for any bounded $w_n$. 
We take the case $\alpha-\gamma<\beta-\delta$ in order to ensure a well-defined steady state and assume an initial Boltzmann distribution 
\begin{equation}
|P_0\rangle = \sum_{n=0}^\infty x^n (1-x) |n\rangle
\end{equation}
where $|n\rangle$ denotes the state with site occupied by $n$ particles and the fugacity $x=e^{-\beta \mu}<1$. 
The steady state is $ x=(\alpha+\delta)/(\beta+\gamma)$ while $x \to 0$ gives the empty site.  By ergodicity this gives the same asymptotic current distribution as any fixed initial particle number.

Explicit computations yield an integral form for the generating function of input current 
\begin{multline}
\langle s | e^{-\tilde{H}_0 t} |P_0 \rangle = \frac{ x-1}{2\pi i} \biggl\{ \oint_{C_1} e^{-\varepsilon(z)t} \frac{1}{(z-1)(z- x)}\, dz \\
+ \oint_{C_2} e^{-\varepsilon(z)t} \frac{ x^{-1} [u_\lambda/v_\lambda - z u_\lambda/(\beta + \gamma)]}{(z-1)[z- x^{-1}u_\lambda/v_\lambda][z-u_\lambda/(\beta + \gamma)]}\, dz  \biggr\} \label{e:intinb}
\end{multline}
with
\begin{equation}
\varepsilon(z)=\alpha+\beta+\gamma+\delta-v_\lambda z- u_\lambda z^{-1}.
\end{equation}
Here, for notational brevity we write
$
u_\lambda \equiv {\alpha e^{-\lambda} +\delta}
$, 
$
v_\lambda \equiv {\beta + \gamma e^\lambda}.
$
and for later use also define the parameter combination
$
\eta=\sqrt{[(\beta+\gamma)^2-\beta\delta-\alpha\gamma]^2-4\alpha\beta\gamma\delta}.
$
The contour $C_1$ ($C_2$) is an anti-clockwise circle of radius $ x+\epsilon$ ($\epsilon$) around the origin of the complex plane with $\epsilon \to 0$. 

In order to extract the large-time behaviour from this integral representation we use a saddle-point method, taking careful account of the contributions from residues when the saddle-point contour is deformed through poles in the integrand. 
This yields changes in behaviour at the values of $\lambda$ given in Table~\ref{t:lam}.  
\begin{table}
\caption{\label{t:lam} Transition values for input current fluctuations in single-site PAZRP}
\begin{largetabular}{ll}
Values of $\lambda$ & Corresponding values of $j$ \\
$e^{\lambda_1} \equiv \frac{\alpha}{\beta+\gamma-\delta}$ &  $j_a \equiv \frac{(\beta+\gamma-\delta)^2-\alpha\gamma}{\beta+\gamma-\delta}$, $j_b \equiv \frac{\beta(\beta+\gamma-\delta)^2-\alpha\gamma\delta}{(\beta+\gamma)(\beta+\gamma-\delta)}$ \\
$e^{\lambda_2} \equiv \frac{(\beta+\gamma)^2-\alpha\gamma-\beta\delta+\eta}{2\gamma\delta}$ & $j_c \equiv -\frac{\eta}{\beta+\gamma}$ \\
$e^{\lambda_3} \equiv \frac{\delta-\beta x^2+\sqrt{(\delta-\beta x^2)^2+4\alpha\gamma x^2}}{2\gamma x^2} $ & $j_d\equiv \frac{-(\delta-\beta x^2)}{ x}$ \\
$e^{\lambda_4} \equiv \frac{\beta(1- x)+\gamma}{\gamma  x} $ &  $j_e \equiv \frac{\alpha\beta\gamma x^2-\delta\left[\beta(1- x)+\gamma \right]^2}{ x(\beta+\gamma)\left[\beta(1- x)+\gamma \right]} $, $j_f \equiv \frac{\alpha\gamma-\left[\beta(1- x)+\gamma \right]^2}{\beta(1- x)+\gamma}$ \\
\end{largetabular}
\end{table}
For
\begin{equation}
 x< x_c\equiv\frac{-\eta+(\beta+\gamma)^2-\alpha\gamma+\beta\delta}{2\beta(\beta+\gamma)}
\end{equation}
we find
\begin{equation}
e_0(\lambda)=
\begin{cases}
\alpha(1-e^{-\lambda})+\gamma(1-e^\lambda) &  \lambda < \lambda_1 \\
\alpha+\delta-\frac{u_\lambda v_\lambda}{\beta+\gamma} & \lambda_1<\lambda<\lambda_2 \\
\alpha+\beta+\gamma+\delta-2\sqrt{u_\lambda v_\lambda} & \lambda_2<\lambda<\lambda_3 \\
\alpha+\beta+\gamma+\delta-v_\lambda x - u_\lambda x^{-1} & \lambda_3<\lambda
\end{cases}
\end{equation}
whereas for $ x> x_c$ we get
\begin{equation}
e_0(\lambda)= 
\begin{cases}
\alpha(1-e^{-\lambda})+\gamma(1-e^\lambda) &    \lambda < \lambda_1 \\
\alpha+\delta-\frac{u_\lambda v_\lambda}{\beta+\gamma}  &  \lambda_1 < \lambda < \lambda_4 \\
\alpha+\beta+\gamma+\delta-v_\lambda  x - u_\lambda  x^{-1} & \lambda_4 < \lambda. 
\end{cases}
\end{equation}
We note that the form of $e_0(\lambda)$ seen in the regime $\lambda_1<\lambda<\lambda_3$ ($\lambda_4$) is the groundstate eigenvalue of $\tilde{H}_0$~\cite{Me05}. At $\lambda=\lambda_2$ the spectrum of $\tilde{H}_0$ 
becomes gapless. 
The changes at $\lambda_1$ and $\lambda_3$ ($\lambda_4$) correspond to the divergence of $\langle s | \tilde{0} \rangle$ and $\langle \tilde{0} | P_0 \rangle$ respectively.  One immediately sees that the symmetry relation~\eqref{e:GCFTe} is only obeyed for a limited range of $\lambda$.  

\begin{figure}
\psfrag{j }[Cr][Cl]{$j\quad$}
\psfrag{m }[Tc][Bc]{$ x$}
\psfrag{III}[][]{III}
\psfrag{I}[][]{I}
\psfrag{II}[][]{II}
\psfrag{VI}[][]{VI}
\psfrag{V}[][]{V}
\psfrag{IV}[][]{~~~~~~IV}
\psfrag{0}[Tc][Tc]{\scriptsize{0}}
\psfrag{0.1}[Tc][Tc]{\scriptsize{0.1}}
\psfrag{0.2}[Tc][Tc]{\scriptsize{0.2}}
\psfrag{0.3}[Tc][Tc]{\scriptsize{0.3}}
\psfrag{0.4}[Tc][Tc]{\scriptsize{0.4}}
\psfrag{0.5}[Tc][Tc]{\scriptsize{0.5}}
\psfrag{0.6}[Tc][Tc]{\scriptsize{0.6}}
\psfrag{0.7}[Tc][Tc]{\scriptsize{0.7}}
\psfrag{0.8}[Tc][Tc]{\scriptsize{0.8}}
\psfrag{0.9}[Tc][Tc]{\scriptsize{0.9}}
\psfrag{1}[Tc][Tc]{\scriptsize{1.0}}
\psfrag{-0.5s}[Cr][Cr]{\scriptsize{-0.5 }}
\psfrag{-0.4s}[Cr][Cr]{\scriptsize{-0.4 }}
\psfrag{-0.3s}[Cr][Cr]{\scriptsize{-0.3 }}
\psfrag{-0.2s}[Cr][Cr]{\scriptsize{-0.2 }}
\psfrag{-0.1s}[Cr][Cr]{\scriptsize{-0.1 }}
\psfrag{0s}[Cr][Cr]{\scriptsize{0 }}
\psfrag{0.3s}[Cr][Cr]{\scriptsize{0.3 }}
\psfrag{0.2s}[Cr][Cr]{\scriptsize{0.2 }}
\psfrag{0.1s}[Cr][Cr]{\scriptsize{0.1 }}
\onefigure[width=0.48\columnwidth]{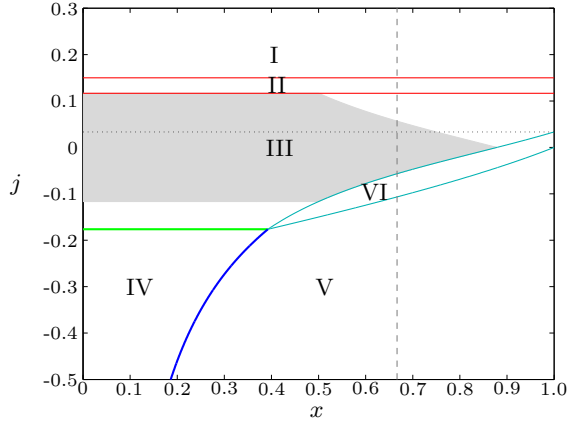}
\caption{Phase diagram for input current large deviations.  Single-site PAZRP with $w_n=1$, $\alpha=0.1$, $\beta=0.2$, $\gamma=0.1$, $\delta=0.1$.  Dotted horizontal line shows mean steady-state current, dashed vertical line denotes steady-state initial condition.  The symmetry~\eqref{e:GCFT} is obeyed in the shaded area inside III.
}
\label{f:pdj}
\end{figure}

Via Legendre transformation we obtain the large deviation behaviour of $j_0=J_0/t$.  The resulting ``phase diagram'' is shown in Fig.~\ref{f:pdj} where
$\hat{e}_0(j)$ has the following forms in the different regions:
\begin{equation}
\hat{e}_0(j)= 
\begin{cases}
f_j(\alpha,\gamma) & \text{I} \\
g_j\!\!\left(\frac{(\alpha-\beta-\gamma+\delta)(\beta-\delta)}{\beta+\gamma-\delta},\frac{\beta+\gamma-\delta}{\alpha}\right) &  \text{II} \\
f_j\!\!\left(\frac{\alpha\beta}{\beta+\gamma},\frac{\gamma\delta}{\beta+\gamma}\right) & \text{III} \\
f_j(\alpha,\gamma)+f_j(\beta,\delta) & \text{IV} \\
f_j(\alpha,\gamma)+g_j(\beta(1- x)+\delta(1- x^{-1}), x) & \text{V} \\
g_j\!\!\left(\frac{(1- x)\left\{\alpha\beta x-\delta\left[\beta(1- x)+\gamma\right]\right\}}{ x\left[\beta(1- x)+\gamma\right]},\frac{\gamma x}{\beta(1- x)+\gamma} \right) & \text{VI} \\
\end{cases} \label{e:ejres} \\
\end{equation}
with
\begin{align}
f_j(a,b)&=a+b-\sqrt{j^2+4ab}+j \ln \frac{j+\sqrt{j^2+4ab}}{2a} \\
g_j(a,b)&=a+j \ln b.
\end{align}
The function $f_j(a,b)$ is the ``random walk'' current distribution of a single bond with Poissonian jumps of rate $a$ to the right and $b$ to the left.  The straightline function $g_j(a,b)$ gives an exponential decay of $p_0(j,t)$ with increasing $j$.  We now give some brief remarks on the physical interpretation of these behaviours. 

In region III, the current across the input bond is dependent on the current across the output bond, resulting in a distribution with mean $(\alpha\beta-\gamma\delta)/(\beta+\gamma)$ and diffusion constant $(\alpha\beta+\gamma\delta)/(\beta+\gamma)$.   In IV ($j$ large and negative) there is a temporary build-up of particles on the site (an ``instantaneous condensate''\cite{Me05}) and so to see $j_0=j$, requires a current of $j$ across both bonds independently.  In I ($j$ large and positive) the piling-up of particles on the site means the input bond does not feel the presence of the output bond.  The $ x$-dependence in region V arises from the possibility of an arbitrarily large initial occupation.  II and VI are transition regimes involving linear combinations of two different behaviours.  They correspond to values of $\lambda$ where $e_0(\lambda)$ has a discontinuous derivative (cf. a first order phase transition).  Analogous results for $e_1(\lambda)$ and $\hat{e}_1(j)$ which characterize the distribution of outgoing current are obtained by the replacements $\alpha \leftrightarrow \delta$, $\beta \leftrightarrow \gamma$, $p \leftrightarrow q$, $\lambda \leftrightarrow -\lambda$, $j \leftrightarrow -j$.  
\begin{figure}
\psfrag{ 0}[Tc][Tc]{\scriptsize{0}}
\psfrag{ 0.05}[Tc][Tc]{\scriptsize{0.05}}
\psfrag{ 0.1}[Tc][Tc]{\scriptsize{0.10}}
\psfrag{ 0.15}[Tc][Tc]{\scriptsize{0.15}}
\psfrag{ 0.2}[Tc][Tc]{\scriptsize{0.20}}
\psfrag{ 0.25}[Tc][Tc]{\scriptsize{0.25}}
\psfrag{ 0.3}[Tc][Tc]{\scriptsize{0.30}}
\psfrag{ 0.35}[Tc][Tc]{\scriptsize{0.35}}
\psfrag{ 0.4}[Tc][Tc]{\scriptsize{0.40}}
\psfrag{-0.05s}[Cr][Cr]{\scriptsize{-0.05}}
\psfrag{-0.1s}[Cr][Cr]{\scriptsize{-0.10}}
\psfrag{ 0s}[Cr][Cr]{\scriptsize{0}}
\psfrag{ 0.05s}[Cr][Cr]{\scriptsize{0.05}}
\psfrag{ 0.1s}[Cr][Cr]{\scriptsize{0.10}}
\psfrag{ 0.15s}[Cr][Cr]{\scriptsize{0.15}}
\psfrag{ 0.2s}[Cr][Cr]{\scriptsize{0.20}}
\psfrag{ 0.25s}[Cr][Cr]{\scriptsize{0.25}}
\psfrag{ 0.3s}[Cr][Cr]{\scriptsize{0.30}}
\psfrag{j}[Tc][Bc]{$j$}
\psfrag{e}[Bc][Tc]{$\hat{e}(-j)-\hat{e}(j)$}
\psfrag{x*log(a*b/(c*d))}[Cl][Cl]{\scriptsize{GC symmetry}}
\psfrag{jin(-x)-jin(x)}[Cl][Cl]{\scriptsize{prediction of~\eqref{e:ejres} for input}}
\psfrag{jout(-x)-jout(x)}[Cl][Cl]{\scriptsize{prediction for output}}
\psfrag{jinmu(-x)-jinmu(x)}[Cl][Cl]{\scriptsize{prediction of~\eqref{e:ejres} for input}}
\psfrag{joutmu(-x)-joutmu(x)}[Cl][Cl]{\scriptsize{prediction for output}}
\psfrag{"GCFT_50led_0GC.dat"}[Cl][Cl]{\scriptsize{in, $t=50$}}
\psfrag{"GCFT_100led_0GC.dat"}[Cl][Cl]{\scriptsize{in, $t=100$}}
\psfrag{"GCFT_200led_0GC.dat"}[Cl][Cl]{\scriptsize{in, $t=200$}}
\psfrag{"GCFT_50led_1GC.dat"}[Cl][Cl]{\scriptsize{out, $t=50$}}
\psfrag{"GCFT_100led_1GC.dat"}[Cl][Cl]{\scriptsize{out, $t=100$}}
\psfrag{"GCFT_200led_1GC.dat"}[Cl][Cl]{\scriptsize{out, $t=200$}}
\psfrag{"GCFTss2_50led_0GC.dat"}[Cl][Cl]{\scriptsize{in, $t=50$}}
\psfrag{"GCFTss2_100led_0GC.dat"}[Cl][Cl]{\scriptsize{in, $t=100$}}
\psfrag{"GCFTss2_200led_0GC.dat"}[Cl][Cl]{\scriptsize{in, $t=200$}}
\psfrag{"GCFTss2_50led_1GC.dat"}[Cl][Cl]{\scriptsize{out, $t=50$}}
\psfrag{"GCFTss2_100led_1GC.dat"}[Cl][Cl]{\scriptsize{out, $t=100$}}
\psfrag{"GCFTss2_200led_1GC.dat"}[Cl][Cl]{\scriptsize{out, $t=200$}}
\psfrag{(a)}{(a)}
\psfrag{(b)}{(b)}
\twoimages[width=0.48\columnwidth]{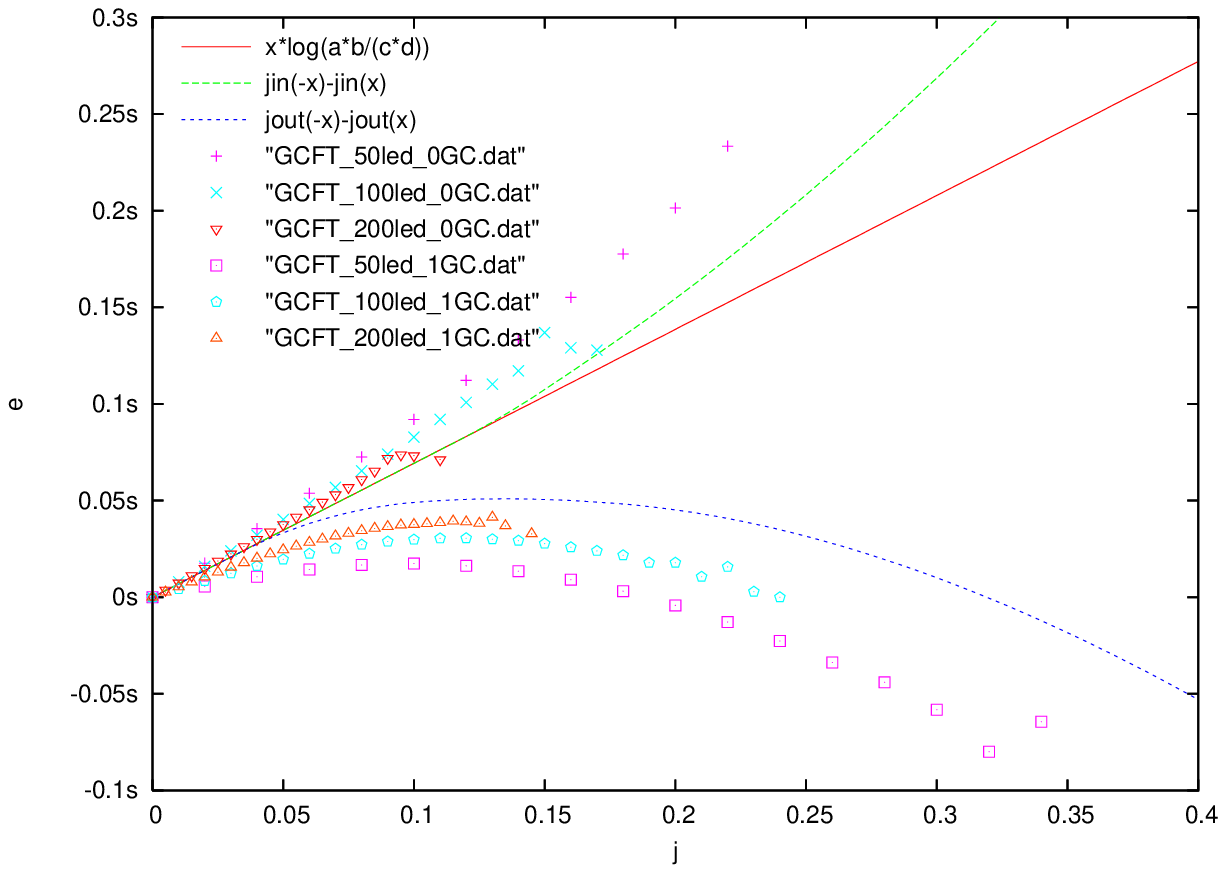}{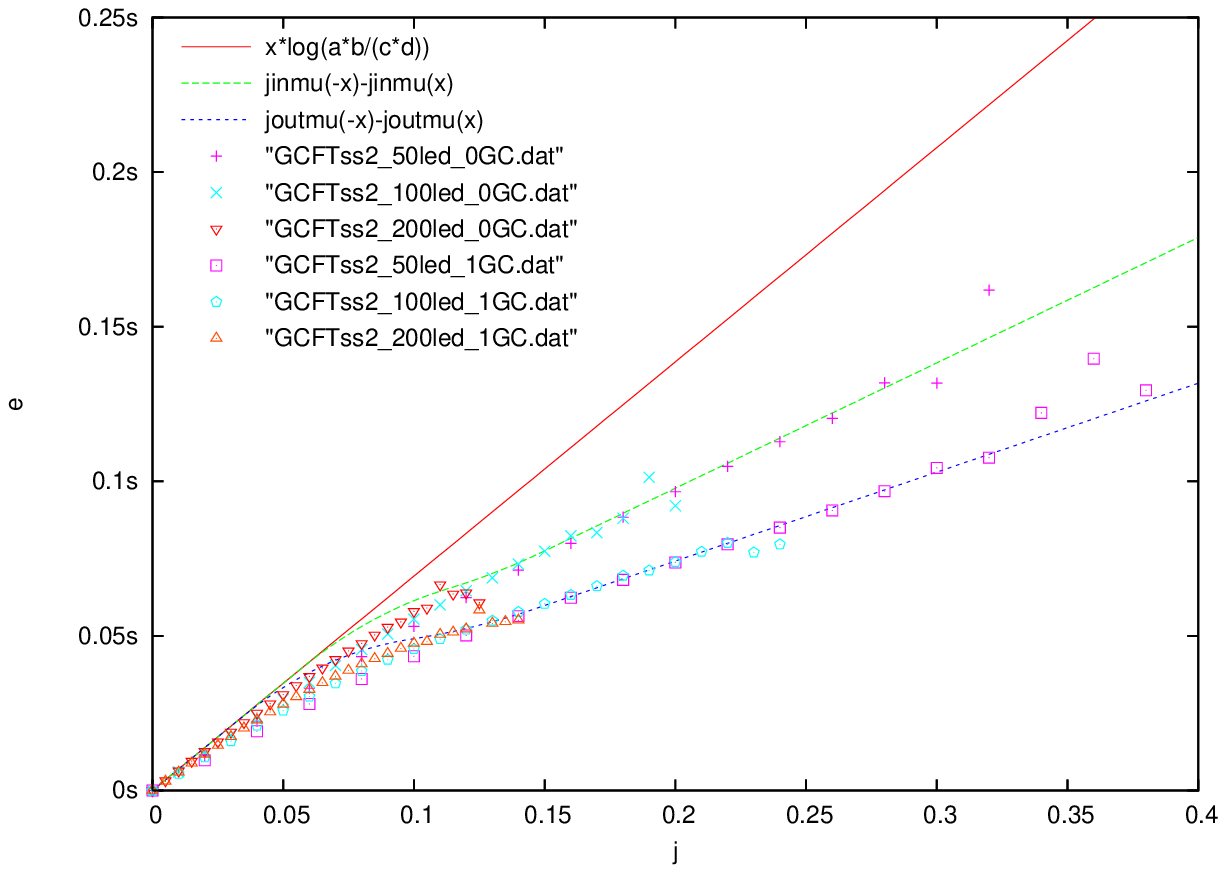}
\caption{Theory (lines) and simulation (points) for $\log[p(j,t)/p(-j,t)]$.  Parameters of Fig~\ref{f:pdj} with initial conditions: (a) $ x=0$ (empty site), (b) $ x=2/3$ (steady state).}
\label{f:GCFT}
\end{figure}

The GC symmetry states that, $\hat{e}(-j)-\hat{e}(j)$ should be a straight line (of slope $\log[(\alpha\beta)/(\gamma\delta)]$ in this single-site case) but the results~\eqref{e:ejres} imply that this only holds for small $j$ (specifically in the shaded region of Fig.~\ref{f:pdj}).  In Fig.~\ref{f:GCFT} we test this prediction against simulation for both input and output bonds.  
The Monte Carlo simulation results were obtained using an efficient event-driven (continuous time) algorithm; for steady-state results the number of histories with each initial occupation was weighted according to the known steady-state distribution~\cite{Levine04c}.  For increasing measurement times the simulation data converges towards the long-time limits predicted by our theory rather than the straight line predicted by GC symmetry. 

Unfortunately, since for increasing times it becomes exponentially more unlikely to measure a current fluctuation away from the mean, it is difficult to get long-time simulation data for a large range of $j$.  
A further check is provided by numerical evaluation of the integral~\eqref{e:intinb} followed by numerical Fourier transform to give the finite time distribution of $p(j,t)$---for small $t$ this gives excellent agreement with the simulation data; for larger $t$ the integrals converge too slowly for the method to be useful.

\begin{figure}
\psfrag{ 0}[Tc][Tc]{\scriptsize{0}}
\psfrag{ 0.02}[Tc][Tc]{\scriptsize{0.02}}
\psfrag{ 0.04}[Tc][Tc]{\scriptsize{0.04}}
\psfrag{ 0.06}[Tc][Tc]{\scriptsize{0.06}}
\psfrag{ 0.08}[Tc][Tc]{\scriptsize{0.08}}
\psfrag{ 0.1}[Tc][Tc]{\scriptsize{0.1}}
\psfrag{ 0.12}[Tc][Tc]{\scriptsize{0.12}}
\psfrag{ 0.14}[Tc][Tc]{\scriptsize{0.14}}
\psfrag{ 0s}[Cr][Cr]{\scriptsize{0}}
\psfrag{ 0.02s}[Cr][Cr]{\scriptsize{0.02}}
\psfrag{ 0.04s}[Cr][Cr]{\scriptsize{0.04}}
\psfrag{ 0.06s}[Cr][Cr]{\scriptsize{0.06}}
\psfrag{ 0.08s}[Cr][Cr]{\scriptsize{0.08}}
\psfrag{ 0.1s}[Cr][Cr]{\scriptsize{0.10}}
\psfrag{ 0.12s}[Cr][Cr]{\scriptsize{0.12}}
\psfrag{-0.02s}[Cr][Cr]{\scriptsize{-0.02}}
\psfrag{-0.04s}[Cr][Cr]{\scriptsize{-0.04}}
\psfrag{j}[Tc][Bc]{$j$}
\psfrag{e }[Bc][Tc]{$\hat{e}(-j)-\hat{e}(j)$}
\psfrag{x*log\(\(a*b\)*\(\(p/q\)**3\)/\(c*d\)\)}[Cl][Cl]{\scriptsize{GC symmetry}}
\psfrag{"w1bnd2_200l_0GC.dat"}[Cl][Cl]{\scriptsize{0th bond}}
\psfrag{"w1bnd2_200l_1GC.dat"}[Cl][Cl]{\scriptsize{1st bond}}
\psfrag{"w1bnd2_200l_2GC.dat"}[Cl][Cl]{\scriptsize{2nd bond}}
\psfrag{"w1bnd2_200l_3GC.dat"}[Cl][Cl]{\scriptsize{3rd bond}}
\psfrag{"w1bnd2_200l_4GC.dat"}[Cl][Cl]{\scriptsize{4th bond}}
\onefigure[width=0.45\columnwidth]{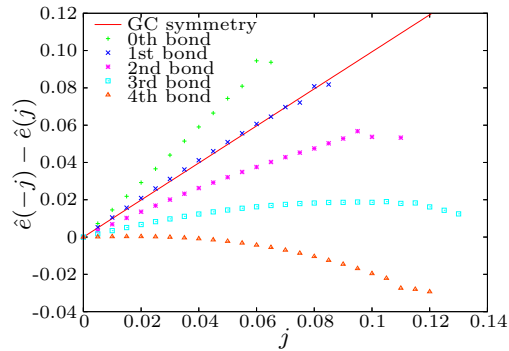}
\caption{Simulation results for $\log[p(j,t)/p(-j,t]$ in four-site PAZRP with $w_n=1-0.5/n$, $\alpha=0.1$, $\beta=0.2$, $\gamma=0.1$, $\delta=0.1$, $p=0.525$, $q=0.475$ and $ x=0$. Points show data for five bonds at $t=200$. 
} 
\label{f:bnd}
\end{figure}

We now turn to numerical results for a larger system with a different choice of bounded $w_n$, see Fig.~\ref{f:bnd}.
In the finite-time simulation regime one again sees indications of violation of GC symmetry with bond-dependent form.
Physically, we argue that the inhomogeneity of the fluctuations across the two different bonds in the single-site PAZRP and the associated violation of the GC symmetry is a result of the temporary build-up of particles on the site.  In general, this possibility is expected to occur in any open-boundary zero-range process with $\lim_{n\to\infty} w_n$ finite (even when the boundary parameters are chosen so that there is a well-defined steady state, i.e., no permanent condensation).  


Mathematically, the observed breakdown of the GC symmetry results from the divergence of $\langle s | \tilde{0} \rangle$ and $\langle \tilde{0} | P_0 \rangle$.  For models where the number of particle configurations $N$ is limited, these quantities are finite and the relation~\eqref{e:GCFT} holds for any initial state.  However, the limit $N \to \infty$ does not necessarily commute with the $t\to\infty$ limit taken (implicitly) in~\eqref{e:GCFT} and (explicitly) in~\eqref{e:e_l}.   This non-commutation of limits leads in some cases to the violation of~\eqref{e:GCFT} \emph{even for steady-state initial conditions}.  This and the initial state dependence (due to non-commuting timescales) are the main issues highlighted by our work.  

We now give a more general explanation of this GC breakdown and highlight some connections to previous works.  Firstly, consider the observed bond dependence.  For systems with bounded state space, currents across different bonds differ by finite boundary terms which vanish in the long-time limit so any combination of currents has the same large deviation behaviour.  In contrast, for systems with unbounded state space, current fluctuations can be spatially inhomogeneous and the boundary terms non-vanishing.  However, there is always a specific weighted sum of currents for which these boundary terms cancel, giving an action functional analogous to heat production (see~\cite{Lebowitz99}).  For the choice $w_n=1$ this is
$
W=2\sum_{l=0}^L E_l J_l
$
where $E_l$ is the effective field across each bond, e.g., for the single-site PAZRP we have $e^{2E_0}=\alpha/\gamma$ and $e^{2E_1}=\delta/\beta$. 

However, it can readily be seen that the large deviations of $W$ still do not satisfy the GC symmetry.  This is due to the presence of further non-vanishing boundary terms.  Consider instead the modified action functional (again for $w_n=1$)
\begin{equation}
W'=2\sum_{l=0}^L E_l J_l - \ln \frac{P_0(\{n\}(t))}{P_0(\{n\}(0))}
\end{equation}
where $\{n\}(t)$ represents the configuration of particles at time $t$ and $P_0$ is the initial distribution.  The fluctuations of this quantity do satisfy the relationship~\eqref{e:GCFT} \emph{even for finite times}---this is a statement of the transient fluctuation theorem of Evans and Searles~\cite{Evans94,Searles99} (see also~\cite{Carberry04,Wang05c} for recent experimental tests).  Only for bounded state space (finite potentials) do the boundary terms containing the initial distribution vanish in the long-time limit leading to recovery of the GC symmetry and the steady-state theorem.

Note that if one measures only a single current (e.g., $J_0$ or $J_1$) but starts with an initial distribution corresponding to
detailed balance across that bond, the boundary terms cancel and the GC symmetry~\eqref{e:GCFT} \emph{is} observed.
A particular example is the zero-current case $\alpha\beta=\gamma\delta$ with an initial equilibrium distribution, $ x=(\alpha+\delta)/(\beta+\gamma)$---the current fluctuations across both bonds become symmetric $\hat{e}(j)=\hat{e}(-j)$ as predicted by the GCFT with $E \to 0$.  This also implies the usual Green-Kubo formula and Onsager reciprocity relations~\cite{Lebowitz99}.  For other values of $x$ a breakdown of~\eqref{e:GCFT} is still predicted in the $E \to 0$ limit (despite the system's ergodicity).

An analogous apparent breakdown of the GCFT in models with \emph{deterministic} dynamics and unbounded potentials was discussed by Bonnetto et al.~\cite{Bonetto05}.  They argue for the restoration of the symmetry by removal of the ``unphysical'' singular terms
An earlier study of a model with both deterministic and stochastic forces~\cite{vanZon03,vanZon04} (see~\cite{Garnier05} for experimental realization) found a modified form of heat fluctuation theorem for large fluctuations.  In contrast to~\cite{Bonetto05,vanZon03,vanZon04}, we do not find a constant value for the ratio of probabilities for large forward and backward fluctuations. 

A. R\'akos acknowledges financial support from the Israel Science Foundation. 

\bibliographystyle{phaip}
\bibliography{allref}

\end{document}